\documentclass{elsart}

\usepackage{natbib}

\usepackage{epsfig}

\usepackage{amssymb}

\begin{document}

\begin{frontmatter}



\title{Influence of ionospheric perturbations in {GPS} time and frequency transfer}


\author{Sophie Pireaux,\corauthref{cor}
}
\corauth[cor]{Corresponding author}
\ead{sophie.pireaux@oma.be}

\hspace{0.8cm}\author{Pascale Defraigne, Laurence Wauters,
}
\ead{Pascale.Defraigne@oma.be, Laurence.Wauters@oma.be}

\hspace{0.5cm}\author{Nicolas Bergeot, Quentin Baire, Carine Bruyninx
}
\address{Royal Observatory of Belgium, 
3 Avenue Circulaire, B-1180 Brussels, Belgium}
\ead{Nicolas.Bergeot@oma.be,Quentin.Baire@oma.be,Carine.Bruyninx@oma.be}

\begin{abstract}

The stability of {GPS} time and frequency transfer is limited by the fact that {GPS} signals travel through the ionosphere.
In high precision geodetic time transfer (i.e. based on precise modeling of code and carrier phase {GPS} data), 
the so-called ionosphere-free combination of the code and carrier phase measurements made on the two frequencies 
is used to remove the first-order ionospheric effect. In this paper, we investigate the impact of residual second- 
and third-order ionospheric effects on geodetic time transfer solutions i.e. remote atomic clock comparisons based on GPS measurements, 
using the {ATOMIUM} software developed at the Royal Observatory of Belgium ({ROB}). 
The impact of  third-order ionospheric effects was shown to be negligible, 
while for second-order effects, the tests performed on different time links and at different epochs show a small 
impact of the order of some picoseconds, on a quiet day, and up to more than 10 picoseconds in case of high ionospheric activity. 
The geomagnetic storm of the 30th October 2003 is used to illustrate how space weather products are relevant to understand perturbations in geodetic time and frequency transfer.

\end{abstract}


\end{frontmatter}

\parindent=0.5 cm


\section{Introduction}
\label{Section_Introduction}

Time and frequency transfer ({TFT}) using {GPS} (Global Positioning System, \citep{L04}) or equivalently {GNSS} (Global Navigation Satellite System) satellites consists in comparing two remote atomic clocks to the reference time scale of the {GPS} system (or to another post-processed time scale based on a {GPS} or {GNSS} network). From the differences between these comparisons, one gets the synchronization error between the two remote clocks and its time evolution. \\
{GPS} {TFT} is widely used within the time community, for example for the realization of TAI (Temps Atomique International), the basis of the legal time {UTC} (Universal Time Coordinated), computed by the Bureau International des Poids et Mesures (BIPM). {TFT} is characterized by its very good resolution (1 observation point/30s or possibly 1 point per second) and a high precision and frequency stability thanks to the carrier phases (uncertainty uA of about 0.1 ns).  The present uncertainty in GPS equipment calibration is 5 ns (uncertainty uB -systematic, hence calibration errors- in the {BIPM} circular T). 

The solar activity varies according to the famous {\it 11-year solar cycle}, also called sunspot cycle. An indicator of the solar activity is the number and intensity of solar flare events, as large flares are less frequent than smaller ones and as the frequency of occurrence varies. 
Solar flares are violent explosions in the Sun's atmosphere that can release as much energy as $6 \cdot 10^{25}$ Joules (more details can be found in \citep{B05}). 
The X-rays and UV radiation emitted by the strongest solar flares can affect the Earth's ionosphere, 
modifying the density and the distribution of the electrons within it.\\
Some solar flares give rise to Coronal Mass Ejections ({CME}s), i.e. 
ejections of plasma (mainly electrons and protons) from the solar corona, carrying a magnetic field and travelling at speeds from about 20 km/s to 2700 km/s, with an average speed of about 500 km/s. Some {CME}s reach the Earth as an Interplanetary {CME} ({ICME}) perturbing the Earth's magnetosphere and hence the Earth's ionosphere.
Consequently, solar flares and associated {CME}s strongly influence our terrestrial environment which has an impact on {GPS}/{GNSS} signals.
%

Since ionospheric influence on electromagnetic waves is frequency dependent and since GPS signals are broadcasted in two different frequencies, ionospheric effects are commonly removed through a given combination (named ionosphere-free) of the signals in the two frequencies $f_{1}$ and $f_{2}$. However, it is well known that this combination removes only first-order perturbations, which correspond to about 99.9\% of the total perturbation. The present study aims at evaluating the impact of the remaining part, concentrating on second- and third-order effects. 
While \cite{FDKRV05} and \cite{HJSO07} investigated the higher-order ionospheric impact on GPS receiver position, GPS satellite clock or GPS satellite position estimates, in the present paper, we focus on the  impact on precise time and frequency transfer using GPS signals. \\
Second- and third-order ionospheric terms are therefore implemented in the software {ATOMIUM} \citep{DGB08}, developed at the Royal Observatory of Belgium. {ATOMIUM} is based on a least-square analysis of dual-frequency carrier phase and code measurements and is able to provide clock solutions in Precise-Point-Positioning ({PPP}) as well as in single-difference (also called Common-View, {CV}) mode. 

The present paper is organized as follows. 
The next section recalls the principles of {GPS} {TFT} and the ionosphere-free analysis in Precise-Point-Positioning or Common-View mode. 
In Section \ref{Section_ATOMIUM}, the ionosphere-free analysis, as implemented in the {ATOMIUM} software, is reviewed. 
In Section \ref{Section_IonoCorr}, the selected method used to implement higher-order ionospheric corrections in the {ATOMIUM} ionosphere-free analysis is described. Our corresponding results are presented in Section \ref{Section_Results}, in terms of ionospheric delays of second and third orders compared to first-order ionospheric effects, and then in terms of the impact of higher-order ionospheric delays in the receiver clock solution computed with {ATOMIUM}. 
In Section \ref{Section_Space_weather}, we put our results in perspective with some investigations on related solar flare events and K-index considerations. Conclusions are finally presented in Section \ref{Section_Conclusions}.\\
Numerical values in the equations of this paper are provided in SI units.

\section{{GPS time and frequency transfer}}
\label{Section_GPS_TFT}
High precision geodetic {GPS} receivers can lock their internal oscillator on an external frequency, given by a stable atomic clock. The {GPS} measurements are then based on the clock frequency. Using post-processed satellite orbits and satellite clock products computed by the International {GNSS} Service ({IGS}), one can deduce the synchronization error between the external clock and either the {GPS} time scale or the reference time scale of the {IGS}, named {IGST}. The use of GPS measurements made in two stations $p$ and $q$ gives then access to the synchronization error between the two atomic clocks in these stations $p$ and $q$. 

For a station $p$ or similarly $q$, the {GPS} measurements, relative to observed satellite $i$, on the signal code $P_{k}$ and phase $L_{k}$, at frequency $k$ ($1$ for $f_{1}=1575.42$ $MHz$ or $2$ for $f_{2}=1227.6$  $MHz$) with corresponding wavelength  $\lambda_{k}$, can be written in length units as
\begin{eqnarray}
\hspace{-0.5 cm}
\left( P_{k=1,2}\right) _{p}^{i} &=&\rho _{p}^{i}+c\Delta t_{p}+c\Delta \tau^{i}
+\Delta r^{i}_{p}+\left(\varepsilon _{P_k}\right) _{p}^{i}
+\left( +I1_{k}+2\cdot I2_{k}+3\cdot I3_{k}\right) _{p}^{i} 
\nonumber\\
\hspace{-0.5 cm}
\left( L_{k=1,2}\right) _{p}^{i} &=&\rho _{p}^{i}+c\Delta t_{p}+c\Delta \tau^{i}
+\Delta r^{i}_{p}+N_{p}^{i}\lambda _{k}+\left(\varepsilon _{L_k}\right) _{p}^{i}
+\left(-I1_{k}-I2_{k}-I3_{k}\right) _{p}^{i}
\nonumber\\
  \overfullrule 5pt
  \mathindent\linewidth\relax
  \advance\mathindent-259pt
\label{eqGPSmeasurementsLP12}
\end{eqnarray}
following \citet{BH93} who considered separately the different orders of the ionosphere impact on GPS code and phase measurements. 
There,  $\rho^{i}_{p}$ is the geometric distance $i$-$p$ ;  $\Delta t_{p}$ is the station clock synchronization error; $\Delta\tau^{i}$ is the satellite clock synchronization error; $\Delta r^{i}_{p}$ is the tropospheric path delay for path $i$-$p$; $I1_{k}$ $(\propto 1/f_k^2)$, $I2_{k}$ $(\propto 1/f_k^3)$ and $I3_{k}$ $(\propto 1/f_k^4)$ are the first-, second- and third-order ionospheric delays on frequency $k$; $N^{i}_{p}$ are the phase ambiguities; $\left(\varepsilon _{P_k}\right) _{p}^{i}$ and  $\left(\varepsilon _{L_k}\right) _{p}^{i}$ are the error terms in code $P$ and phase $L$, respectively, containing noise such as unmodeled multipath, hardware delays (biases from the electronic of the satellite or the station receiver) on the propagation of the modulation/carrier of the signal. 
In their most general form, the different orders of ionospheric delays ($I1$ ,$I2$, $I3$) are expressed as integrals along the true path of the signal (including a bending, which is function of the signal frequency, in the dispersive ionosphere), in terms of the signal frequency, of the local electron density in the ionosphere and of the local geomagnetic field along the trajectory \citep{BH93}. \\
When a dual frequency GPS receiver is available at station $p$, the so-called ionosphere-free combination ($k=3$) is used. This combination is defined as \citep{L04}:
\begin{eqnarray}
P_{3} &\equiv&\frac{f_{1}^{2}}{\left( f_{1}^{2}-f_{2}^{2}\right) }P_{1}-\frac{%
f_{2}^{2}}{\left( f_{1}^{2}-f_{2}^{2}\right) }P_{2} 
\nonumber \\
L_{3} &\equiv&\frac{f_{1}^{2}}{\left( f_{1}^{2}-f_{2}^{2}\right) }L_{1}-\frac{%
f_{2}^{2}}{\left( f_{1}^{2}-f_{2}^{2}\right) }L_{2}
  \overfullrule 5pt
  \mathindent\linewidth\relax
  \advance\mathindent-259pt
\label{eqIonoFreeCombi}
\end{eqnarray}
with $f_{1}$ and $f_{2}$ the two {GPS} carrier frequencies. This combination removes, from the {GPS} signal, the first-order ionospheric effect, $I1$, since the latter is proportional to the inverse of the square frequency.
The corresponding ionosphere-free observation equations therefore do not contain any first-order ionospheric term, but contain new factors for second- and third-order ionospheric effects \citep{FDKRV05} with respect to previous Equation \ref{eqGPSmeasurementsLP12}:
\begin{eqnarray}
\left( P_{3}\right) _{p}^{i} &=&\rho _{p}^{i}+c\Delta t_{p}+c\Delta \tau
^{i}+\Delta r^{i}_{p}+\left(\varepsilon _{P_3}\right) _{p}^{i}+\left( +2\cdot I2_{3}-3\cdot I3_{3}\right)
_{p}^{i} 
\nonumber \\
\left( L_{3}\right) _{p}^{i} &=&\rho _{p}^{i}+c\Delta t_{p}+c\Delta \tau
^{i}+\Delta r^{i}_{p}+N_{p}^{i}\lambda _{3}+\left(\varepsilon _{L_3}\right) _{p}^{i}+\left(
-I2_{3}+I3_{3}\right) _{p}^{i}
  \overfullrule 5pt
  \mathindent\linewidth\relax
  \advance\mathindent-259pt
\label{eqGPSmeasurementsLP3}
\end{eqnarray}
Note that about 99.9\% \citep{H08} of ionospheric perturbations are removed with $I1$ in the so-called ionosphere-free combination. Note also that while the first order has the same magnitude on {GPS} phase and code measurements (but with opposite sign), the impact of second- and third-order effects is larger on code than on phase observations (twice for $I2$, three times for $I3$). 
The GPS observation equations given by \ref{eqGPSmeasurementsLP3} are directly used in Precise Point Positioning. 
The clock solution obtained is the synchronization error between the receiver clock and the {GPS} or {IGS} Time scale. 

For Common-View analysis, one uses the single differences between simultaneous observations of a same satellite $i$ in two remote stations $p$ and $q$ in order to determine directly the synchronization error between the two remote clocks: the observation equations for receivers $p$ and $q$ with satellite $i$ are subtracted. This single difference removes the satellite clock bias in the {GPS} signal, assuming that the nominal times of observation of the satellite by the two stations are the same. 
When forming ionosphere-free combinations, the single-difference code and carrier phase equations are: 
\begin{eqnarray}
\left( P_{3}\right) _{pq}^{i} &=&\rho _{pq}^{i}+c\Delta
t_{pq}+\Delta r^{i}_{pq}+\left(\varepsilon _{P_3}\right) _{pq}^{i}+\left( +2\cdot I2_{3}-3\cdot I3_{3}\right)
_{pq}^{i} 
\nonumber \\
\left( L_{3}\right) _{pq}^{i} &=&\rho _{pq}^{i}+c\Delta
t_{pq}+\Delta r^{i}_{pq}+N_{pq}^{i}\lambda _{3}+\left(\varepsilon _{L_3}\right) _{pq}^{i}+\left(
-I2_{3}+I3_{3}\right) _{pq}^{i}
  \overfullrule 5pt
  \mathindent\linewidth\relax
  \advance\mathindent-259pt
\label{eqGPSmeasurementsLP3CV}
\end{eqnarray}
where, for any quantity $X$,
\begin{eqnarray}
X_{pq}\equiv X_{p}-X_{q}
\nonumber
  \overfullrule 5pt
  \mathindent\linewidth\relax
  \advance\mathindent-259pt
\label{eqXdefinition}
\end{eqnarray}

Now, all the terms in above Equations \ref{eqGPSmeasurementsLP12}, \ref{eqGPSmeasurementsLP3} or \ref{eqGPSmeasurementsLP3CV} can be estimated via an inversion procedure using some a priori precise satellite orbits and satellite clock products. This finally provides the solution for either $\Delta t_{p}$ in {PPP}, i.e. the clock synchronization error between the atomic clock connected to the {GPS} receiver and the {GPS} or {IGS} Time scale at each epoch, or $\Delta t_{pq}$ in Common View, i.e. the synchronization error between the remote clocks connected to two {GPS} receivers. In parallel, the station position and tropospheric zenith delays are estimated as a by-product.

\section{The ATOMIUM software}
\label{Section_ATOMIUM}

\begin{figure}[p]
\begin{center}
\includegraphics*[width=12cm,angle=0]{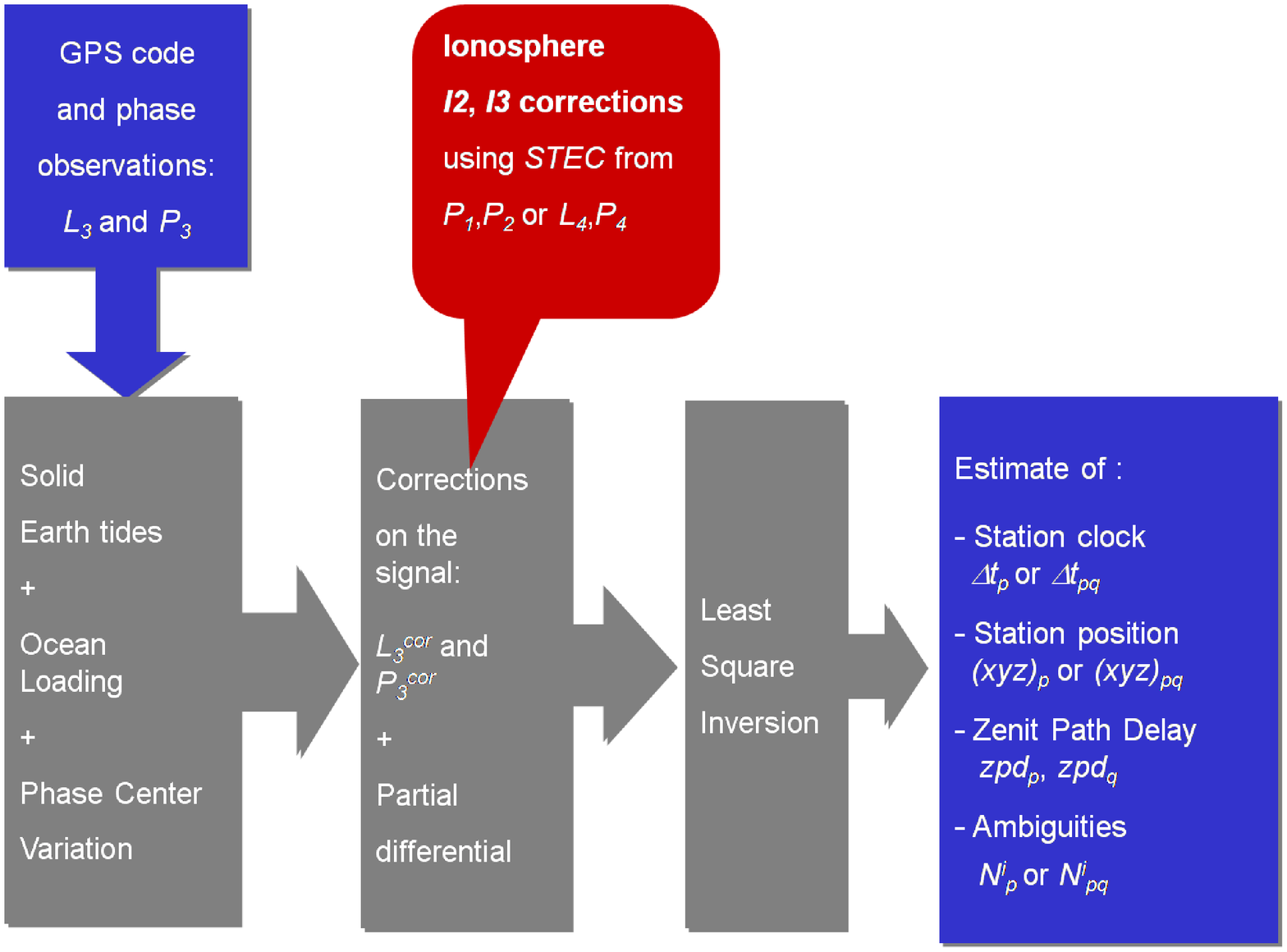}
\end{center}
\caption{ATOMIUM software diagram.}
\label{AtomiumDiagram}
\end{figure}

The present study on ionospheric higher-order perturbations in {TFT} is based on the {ATOMIUM} software \citep{DGB08}, which uses a weighted least-square approach with ionosphere-free combinations of dual-frequency {GPS} code ($P_{3}$) and carrier phase ($L_{3}$) observations. {ATOMIUM} was initially developed to perform {GPS} {PPP}, as described in \citep{KH01}, and later adapted to single differences, or Common View ({CV}), of {GPS} code and carrier phase observations. In the paragraphs below, we describe {ATOMIUM}, following the diagram presented in Figure \ref{AtomiumDiagram}.
First, {GPS} ionosphere-free code and phase combinations are constructed according to Equations \ref{eqIonoFreeCombi} from $L_{1}$, $P_{1}$, $L_{2}$, $P_{2}$ observations read in {RINEX} observation files. By default, the ATOMIUM software uses as a priori the {IGS} products \citep{IGSa09}. IGS satellite clocks (tabulated with a 5-minute interval) are used to obtain $\Delta \tau^{i}$ at the same sampling rate as provided. {IGS} satellite orbits (tabulated with a 15-minute interval) are used to estimate $\rho_{p}^{i}$ (or $\rho_{pq}^{i}$ ) via a 12-points Neville interpolation of the satellite position every 5 minutes.
The station position is corrected for its time variations due to degree 2 and 3 solid Earth tides as recommended by the {IERS} conventions \citep[][Ch. 7]{MP03} and for ocean loading according to the {FES2004} model \citep{LLL04}. The relative (if {GPS} observations were made before {GPS} week 1400) or absolute (if after) elevation (no azimuth) and nadir dependant corrections for receiver and satellite antenna phase center variations are read from the {IGS} atx file available at \citep{IGSb09}.
Prior to the least-square inversion, the computed geometric distance is removed from both phase and code ionosphere-free combinations. Those are also corrected for a relativistic (periodic only) delay and an hydrostatic tropospheric delay. Tropospheric delays are modeled as the sum of a hydrostatic and a wet delay, resulting from the product of a given mapping function and of the corresponding hydrostatic or wet zenith path delay ($zpd$). For the hydrostatic part, we use the Saastamoien a priori model \citep{S72} and the dry Niell mapping function \citep{N96}. For the wet part, we use the wet Niell mapping function \citep{N96} while the wet $zpd$ is estimated as one point every 2 hour, and modeled by linear interpolation between these points.
Carrier phase measurements are further corrected for phase windup \citep{WWHBL93} taking into account satellite attitude and eclipse events.
The implementation of additional higher-order ionospheric corrections on phase and code is done at this level, as corrections applied on the code and phase measurements (see Section \ref{Section_IonoCorr}).

The least-square analysis used in ATOMIUM is detailed in \citep{DGB08}.
As output, {ATOMIUM} provides the station $p$ (or relative $p$-$q$) position for the whole day, the receiver clock $p$ (or relative $p$-$q$) synchronization error every 5 minutes, tropospheric wet zenith path delays $p$ (and $q$) at a given rate (2 hours in our case).

%

\begin{figure}[p]
\begin{center}
\includegraphics*[trim = 0mm 0mm 0mm 0mm,clip,width=6.95cm,angle=0]{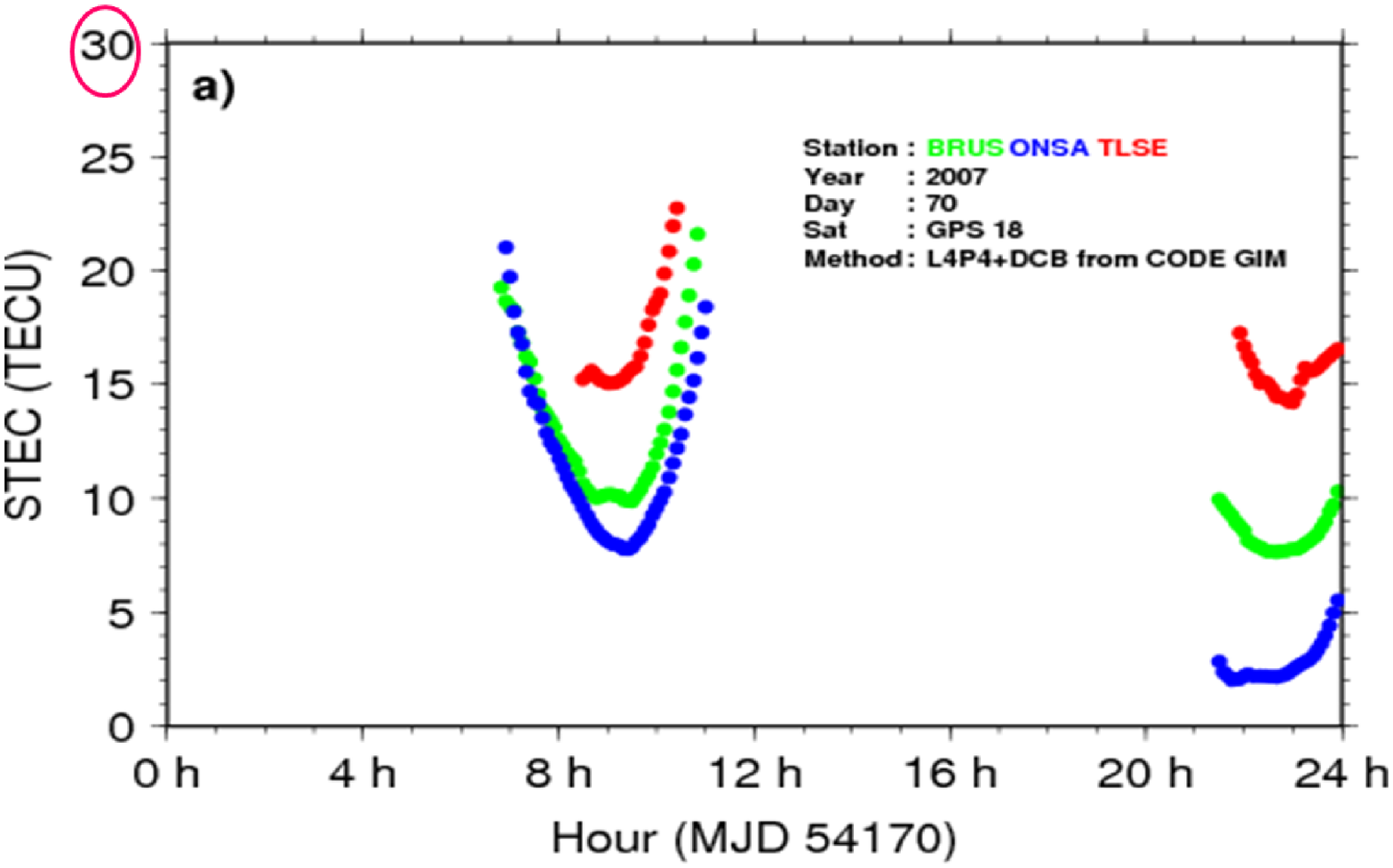}
\includegraphics*[trim = 12mm 0mm 0mm 0mm,clip,width=6.75cm,angle=0]{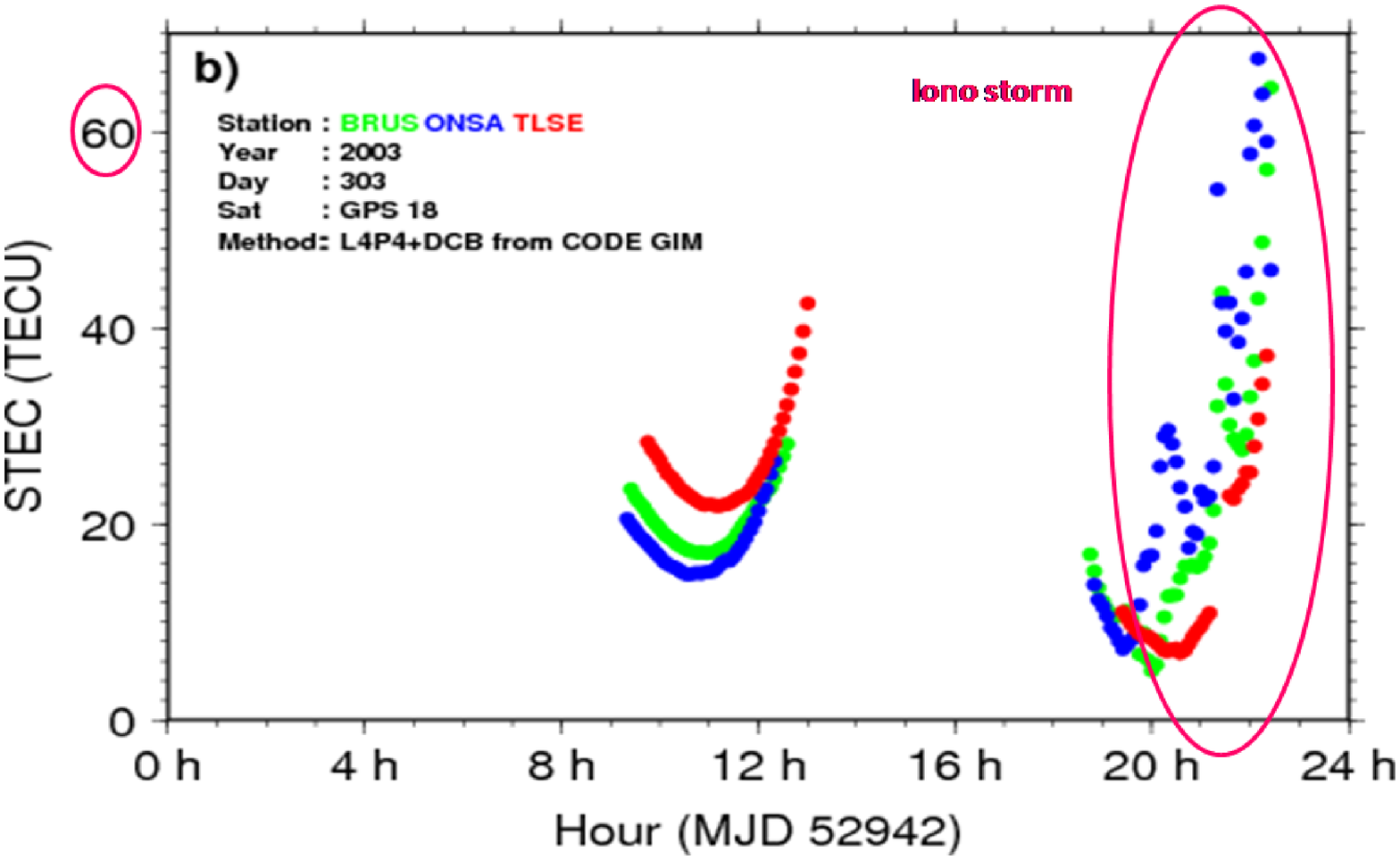}
\end{center}
\caption{{STEC} computed when observing GPS satellite prn 18 with GPS receivers located at Brussels (BRUS), Onsala (ONSA) and Toulouse (TLSE) on an ionosphere-quiet day (11th March 2007), 
Figure a (left), or on a stormy day (30th October 2003), Figure b (right). 
STEC was computed with the ATOMIUM software, using Equation (9) and DCBs read from CODE IONEX files. 
1 {TECU}$=10^{16}$ $e^{-}/m^{2}$.}
\label{STECbrusonsatlse}
\end{figure}

\section{Corrections for ionospheric delays}
\label{Section_IonoCorr}
First-, second- and third-order ionospheric terms are function of the Slant Total Electron Content ({STEC}), 
which is the integrated electron density inside a cylinder column of unit base area between Earth ground and
satellite altitude, along the satellite $i$- station $p$ direction. 
{STEC} is function not only of the satellite elevation or station position, but also of the time of the day, 
of the time of the year, of the solar cycle, and of the particular ionospheric conditions. 
As the $I1$ term contains 99.9\% of the ionospheric perturbations on the GNSS signal, it can be used to estimate STEC. The second- and third-order terms are then computed using this estimated STEC.
The STEC in $I1$ can be determined using the geometry-free combinations of the measurements made on the two {GPS} frequencies, defined as $P_{4}$ and $L_{4}$ \citep{L04}:
\begin{eqnarray}
P_{4} &\equiv&-P_{1}+P_{2}
\nonumber \\
L_{4} &\equiv&L_{1}-L_{2}
  \overfullrule 5pt
  \mathindent\linewidth\relax
  \advance\mathindent-259pt
\label{eqGeometryFreeCombi}
\end{eqnarray}
The quantities ($P_{4}L_{4}$) only contain a given combination of ionospheric delays on the signals of frequency $f_{1}$ and $f_{2}$, plus some constant terms associated with the differential hardware delays (between $f_1$ and $f_2$) in the satellite and in the receiver, and the phase ambiguities.

In the following, for a practical and efficient implementation of higher-order ionospheric terms in ATOMIUM, we work in the no-bending approximation, meaning that the trajectory considered to compute those ionospheric corrections on {GPS} observations is a straight line from satellite $i$ to station $p$ (and $q$). 
Indeed, from analytical estimations based on results of \citet[][formula (31)]{HJ08} and \citet[][formulas (12-13)]{JPM94}, we quantified the bending effects in the dispersive ionosphere on the ionosphere-free combinations of GPS measurements. These effects are due to the excess path length for a curved trajectory and to the difference in effective STEC for the two GPS frequencies. Under extreme conditions, i.e. a highly ionized ionosphere together with very low satellite elevations, ionospheric bending effects might reach the level of the third-order ionosphere correction, which turns out to be in the TFT noise as we will show in the following.\\
We furthermore assume the ionosphere Single Layer Model (SLM) \citep{BH93}, reducing the whole ionosphere layer through which the GPS signal travels to a single effective sheet at a given height with the equivalent electron content (see Figure 3). \\
The above assumptions allow to reduce the general integrals for the ionospheric delays to the easily implemented equations given in the following subsections.

\subsection{First-order ionospheric delays}

The first-order ionospheric effect, is given by \citep{BH93}
\begin{eqnarray}
I1_{k} &=&\alpha 1_{k}\cdot STEC
  \overfullrule 5pt
  \mathindent\linewidth\relax
  \advance\mathindent-259pt
\label{eqFirstOrderIonoEffect12}
\end{eqnarray}
with the factor for {GPS} frequencies 1 and 2 being
\begin{eqnarray}
\alpha 1_{1,2} &=&+\frac{40.3}{f_{1,2}^{2}}
  \overfullrule 5pt
  \mathindent\linewidth\relax
  \advance\mathindent-259pt
\label{eqFirstOrderIonoFactor12}
\end{eqnarray}
This implies that the corresponding factors for the ionosphere-free ($k=3$) or geometry-free ($k=4$) combinations are
\begin{eqnarray}
\alpha 1_{3} &=&0
\nonumber \\
\alpha 1_{4} &=&-40.3\left( \frac{1}{f_{1}^{2}}-\frac{1}{f_{2}^{2}}\right)
  \overfullrule 5pt
  \mathindent\linewidth\relax
  \advance\mathindent-259pt
\label{eqFirstOrderIonoFactor34}
\end{eqnarray}

As announced here above, for each pair of code or phase measurements (on $f_{1}$ and $f_{2}$), the geometry-free combination can be used to compute the {STEC}, which is needed in higher-order ionospheric corrections. Neglecting the $I2$ and $I3$ contributions inducing errors in estimated {STEC} of the order of 0.1 TECU at the most, one gets  \citep{HJSO07}:
\begin{eqnarray}
\hspace{-0.8cm}
\left( STEC\right) _{p}^{i} &=&\frac{1}{\alpha 1_{4}}
\left[ 
\left(L_{4}\right) _{p}^{i}
-\left\langle \left( L_{4}\right) _{p}^{i}-\left(
P_{4}\right) _{p}^{i}\right\rangle _{
\stackrel{arc\ without}{cycle\ slips}
}
-c\cdot DCB_{p}-c\cdot DCB^{i}
\right] 
  \overfullrule 5pt
  \mathindent\linewidth\relax
  \advance\mathindent -259pt
\label{eqSTECarc}
\end{eqnarray}
In the above formula, $P_{1}-P_{2}$ Differential Code Biases ({DCB}) are assumed constant during a day, and we read them from the {CODE} ({IGS} Analysis Center) {IONEX} files; $< >$ means taking the average. \\
Alternatively, {STEC} can be computed using $P_{1}P_{2}$ codes that have first been smoothed with the corresponding phase \citep{DHFM07},
\begin{eqnarray}
\left( STEC\right) _{p}^{i} &=&\frac{1}{\alpha 1_{4}}\left[ \left\{ \left(
P_{2}\right) _{p}^{i}-\left( P_{1}\right) _{p}^{i}\right\} _{
\stackrel{smoothed}{with\ phase}  
}
-c\cdot DCB_{p}-c\cdot DCB^{i}\right]
  \overfullrule 5pt
  \mathindent\linewidth\relax
  \advance\mathindent-259pt
\label{eqSTECsmooth}
\end{eqnarray}
This leads to similar results as those obtained from Equation \ref{eqSTECarc} with respect to the same {DCB} product. \\
Within ATOMIUM, {STEC} is computed, using formula \ref{eqSTECarc}, for each satellite-station pair with a sampling rate of 5 minutes. 
Figure \ref{STECbrusonsatlse} illustrates STEC computed via ATOMIUM for stations {BRUS}, i.e. Brussels, 
at latitude 50$^{\circ}$28' and longitude 4$^{\circ}$12', {TLSE}, i.e. Toulouse at latitude 43$^{\circ}$20' and longitude
 1$^{\circ}$17', {ONSA} i.e. Onsala at latitude 57$^{\circ}$14' and longitude 11$^{\circ}$33'. 
Figure \ref{STECbrusonsatlse}a features a quiet-ionosphere day (around a minimum of solar activity), 11th March 2007, with its normal local-noon diurnal maximum of STEC. 
Figure \ref{STECbrusonsatlse}b shows the ionospheric storm of October 30, 2003 and the STEC sensitivity to perturbed ionospheric conditions as induced by solar activity (see Section \ref{Section_Space_weather}). 
Since ionospheric effects in {GPS} are function of {STEC}, we understand that ionosphere-induced errors will increase in the next few years due to the increasing solar activity associated with the ascending phase of the 24th sunspot cycle (maximum forecast around 2011-2012 depending on the models).

\subsection{Second-order ionospheric delays}

\begin{figure}[b]
\begin{center}
\includegraphics*[width=10cm,angle=0]{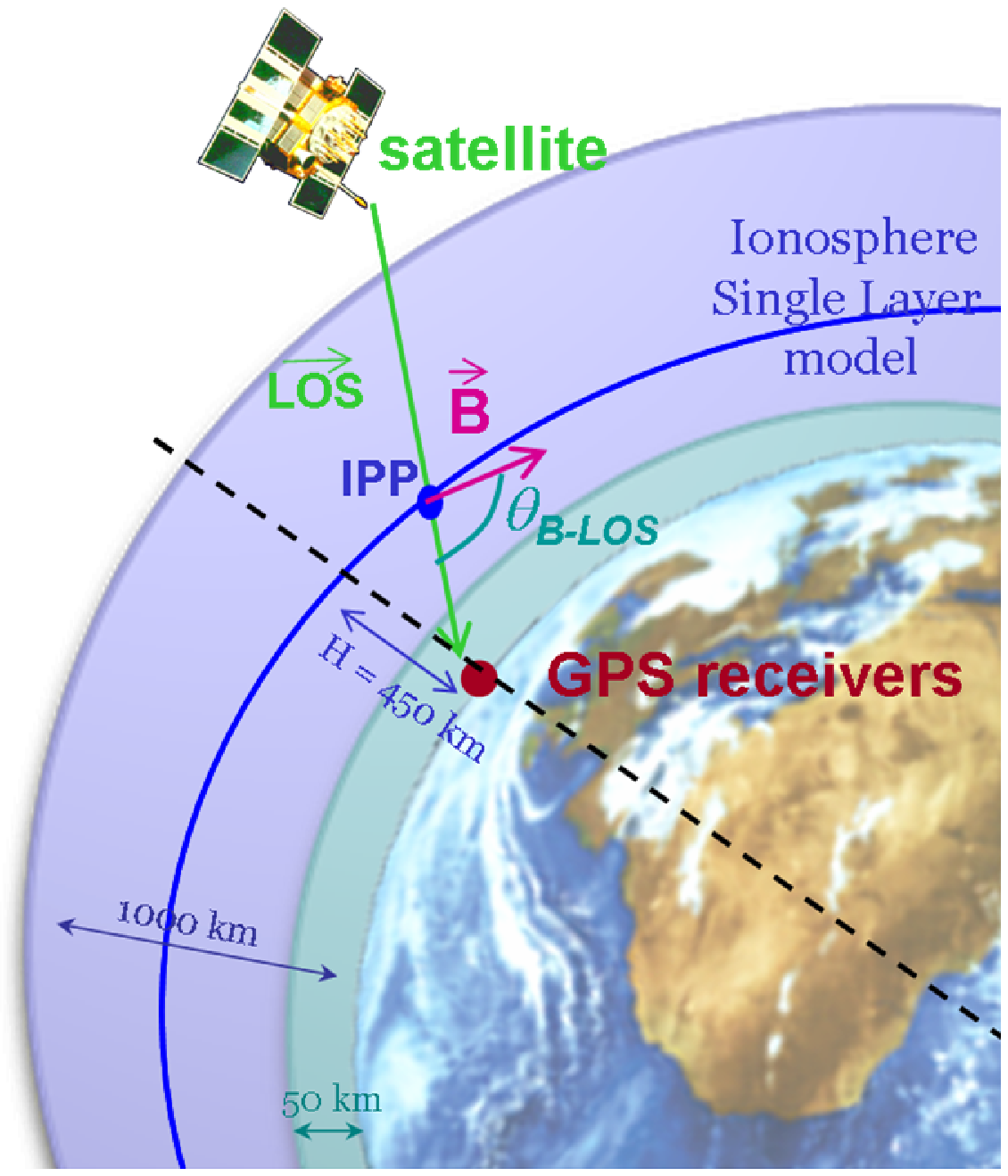}
\end{center}
\caption{The second-order ionospheric effect is not only a function of {STEC}, but it is also function of the angle between the Line Of Sight ({LOS}) and the geomagnetic field $B$ at the Ionosphere Piercing Point ({IPP}), and of the magnitude of $B$ at {IPP}.}
\label{IPP}
\end{figure}

Whereas the magnitude of $I1$ for a given frequency depends solely on {STEC} and is always positive, the magnitude and sign of $I2$ depend on the geomagnetic field $B$ values, on {STEC} and on the $i$-$p$ signal direction via the angle between $B$ and the Line of Sight (LOS), $\theta_{B-LOS}$ (Figure \ref{IPP}). We used the following integrated formula 
\citep{HJSO07}
\begin{eqnarray}
I2_{k}=\alpha 2_{k}\cdot B_{IPP}\cdot \cos \theta _{B-LOS}\cdot STEC
  \overfullrule 5pt
  \mathindent\linewidth\relax
  \advance\mathindent-259pt
\label{eqSndOrderIonoEffect}
\end{eqnarray}
with frequency factors
\begin{eqnarray}
\alpha 2_{3} &=&-\frac{7527\cdot c}{2\cdot f_{1}f_{2}\left(
f_{1}+f_{2}\right) }
  \overfullrule 5pt
  \mathindent\linewidth\relax
  \advance\mathindent-259pt
\label{eqSndOrderIonoFactor3}
\end{eqnarray}
\begin{eqnarray}
\alpha 2_{1,2} &=&-\frac{7527\cdot c}{f_{1,2}^{3}}
  \overfullrule 5pt
  \mathindent\linewidth\relax
  \advance\mathindent-259pt
\label{eqSndOrderIonoFactor12}
\end{eqnarray}
{STEC} is obtained from $L_{4}P_{4}$ (Equation \ref{eqSTECarc}) and $B_{IPP}$ is computed using the accurate International Geomagnetic Reference ({IGR}) model \citep{T05}, as the latter allows to reduce errors in $I2$ up to 60\% with respect to a dipolar model (Hernandez-Pajares et al., 2007, 2008).

\subsection{Third-order ionospheric delays}

In the third-order ionospheric contribution, the magnetic field term can be safely neglected at sub-millimeter error level, leading to the simple formula \citep{FDKRV05}
\begin{eqnarray}
I3_{k}=\alpha 3_{k}\cdot STEC
  \overfullrule 5pt
  \mathindent\linewidth\relax
  \advance\mathindent-259pt
\label{eqThirdOrderIonoEffect}
\end{eqnarray}
with frequency factors also being functions of the electron distribution in the ionosphere:
\begin{eqnarray}
\alpha 3_{1,2} &=&-\frac{2437\cdot N_{\max }\cdot \eta }{f_{1,2}^{4}}
  \overfullrule 5pt
  \mathindent\linewidth\relax
  \advance\mathindent-259pt
\label{eqThirdOrderIonoFactor12}
\end{eqnarray}
\begin{eqnarray}
\alpha 3_{3} &=&-\frac{2437\cdot N_{\max }\cdot \eta }{3\cdot f_{1}^{2}\cdot
f_{2}^{2}}
  \overfullrule 5pt
  \mathindent\linewidth\relax
  \advance\mathindent-259pt
\label{eqThirdOrderIonoFactor3}
\end{eqnarray}
where the shape factor $\eta$ is taken around 0.66 and the peak electron density along the signal propagation path, $N_{max}$, can be determined by a linear interpolation between a typical ionospheric situation and a solar maximum one \citep[][formula 14 corrected according to a private communication from M. Fritsche, November 2008]{FDKRV05}, using, as interpolation coefficients, the numerical values suggested by \citet{BG91}:
\begin{eqnarray}
N_{\max }=\frac{\left[ \left( 20-6\right) \cdot 10^{12}\right] }{\left[
\left( 4.55-1.38\right) \cdot 10^{18}\right] }\cdot \left( VTEC-4.55\cdot
10^{18}\right) +20\cdot 10^{12}
  \overfullrule 5pt
  \mathindent\linewidth\relax
  \advance\mathindent-259pt
\label{eqNmax}
\end{eqnarray}
The Vertical {TEC} ({VTEC}), which is {TEC} along a vertical trajectory below the satellite, is taken as the projection, via the Modified Single Layer Model ionosphere mapping function, of $(STEC)^{i}_{p}$ from Equation \ref{eqSTECarc} with  $\alpha_{MSLM}=0.9782$, $R_{Earth} =6371\cdot 10^{3}$ m, $H=506.7\cdot 10^{3}$ m as in \citep{DHFM07}: 
\begin{eqnarray}
STEC=f_{MSLM}(z)\cdot VTEC
  \overfullrule 5pt
  \mathindent\linewidth\relax
  \advance\mathindent-259pt
\label{eqSTECvsVTEC}
\end{eqnarray}
\begin{eqnarray}
f_{MSLM}(z)\equiv 1/\sqrt{1-\left( \frac{R_{Earth}}{R_{Earth}+H}\right)
^{2}\cdot \cos ^{2}\left( \alpha _{MSLM}\cdot z\right) }
  \overfullrule 5pt
  \mathindent\linewidth\relax
  \advance\mathindent-259pt
\label{eqMappingFunction}
\end{eqnarray}
Finally, using the above equations for the first-, second- and third-order ionospheric effects on {GPS} signal propagation, one finds the orders of magnitude of their impact on the code and carrier phase measurements given in Table \ref{table}.

\section{Quantifying ionospheric effects on TFT}
\label{Section_Results}

The $I2$ and $I3$ corrections computed according the procedure described above were applied to the ionosphere-free combinations $P_{3}$ and $L_{3}$ used in {ATOMIUM}. The present section shows some preliminary results: estimated second- and third-order delays on {GPS} signals (and on combinations of their measurements) and the impact of these on the time and frequency transfer solutions. 

\subsection{Ionospheric delays}

The first results concern the ionospheric 
delays as computed with {ATOMIUM} according to the models detailed in previous section. \\
Firstly, recall that 
the Total Electron Content of the ionosphere reaches its maximal value at local noon, on a normal day (Figure \ref{STECbrusonsatlse}). Since $I1$, $I2$ and $I3$ are proportional to Slant {TEC}, the amplitude of ionospheric perturbations in {GPS} signal reflect this daily variation of {TEC}. \\
Secondly, the diurnal TEC maximum is function of the station latitude, so the ionospheric delays in GPS signals follow accordingly. \\
And finally, for any observed satellite, as the ionospheric thickness crossed by the signal is proportional to the inverse of the sine of the satellite elevation, the {STEC} during one satellite track, as well as the ionospheric delays, takes the shape of a concave curve.\\
Figures \ref{1stOrderIonoPerturbation}, \ref{2ndOrderIonoPerturbation} and \ref{3rdOrderIonoPerturbation} illustrate $I1_{2}$, $I2_{3}$ and $I3_{3}$ respectively on a quiet (left) versus an ionosphere-stormy day, the ionospheric storm of October 30, 2003 (right). The selected station in this illustration is Onsala ({ONSA}). Each point of the curves corresponds to an ionospheric correction on a GPS phase measurement for an observed satellite. 
The first-order ionospheric perturbations in $L_{2}$ can reach about 100 nanoseconds during the storm (Figures \ref{1stOrderIonoPerturbation} a and b), while it is less than 50 nanoseconds in normal times. $I1$ in $L_{1}$ is slightly smaller according to factor $f_{2}^{2}/ f_{1}^{2}$. 
The $I1$ effect is removed from the ionosphere-free combination. \\
The second-order ionospheric perturbation in the ionosphere-free combination (Figure \ref{2ndOrderIonoPerturbation} a) is about 3 to 4 orders of magnitude smaller than the first order in $L_{2}$. $I2_{3}$ can reach about 15 picoseconds during the storm, about the double of its maximum value during a quiet day (Figure \ref{2ndOrderIonoPerturbation}b compared to \ref{2ndOrderIonoPerturbation}a).
We also recall that, in the ionosphere-free combination, the second-order ionospheric perturbation, $I2_{3}$, affects twice more the codes than the phases, as seen in Equation \ref{eqGPSmeasurementsLP3}. \\
Figure \ref{3rdOrderIonoPerturbation} illustrates the third-order ionospheric perturbation which is again an order of magnitude smaller than the second order. Here the effect of the storm is also clear, as the third-order effect in the ionosphere-free combination can reach about 2 picoseconds during the storm (Figure \ref{3rdOrderIonoPerturbation}b), while its maximum value on a non-stormy day is about 0.14 picoseconds (Figure \ref{3rdOrderIonoPerturbation}a). Again, the contribution of $I3_{3}$ is three times more important for codes than for phases, as seen in Equation \ref{eqGPSmeasurementsLP3}, but remains negligible with respect to the present precision of {GPS} time and frequency transfer.

\subsection{Ionospheric impact on receiver clock estimates from a $L_{3}P_{3}$ analysis}

Table \ref{table} and the results presented in the above paragraph illustrate the need to take second-order ionospheric corrections into account in $P$ and $L$ measurements for {TFT}. However, to be coherent, in addition to the $I2$ (and $I3$) correction(s) on {GPS} code and phase data, we should also use satellite orbit and clock products computed with $I2$ (and $I3$) correction(s) in order to estimate the impact of the ionosphere on station clock synchronization errors via {ATOMIUM}. Indeed, \citet{HJSO07} estimated that second-order ionospheric effects in satellite clocks were the largest and could be more than 1 centimeter (i.e. $\sim$ 30 picoseconds);  the same authors mentioned that the second-order ionospheric effects on the satellite position are of the order of several millimeters only, and consist in a global southward shift of the constellation. Current {IGS} products do not take $I2$ nor $I3$ into account,
this is why we present here the impact of our ionospheric corrections on clock solutions via {ATOMIUM} in Common-View mode (Figures 
\ref{Iono2Effect} and \ref{Iono3Effect}), as the satellite clock is eliminated in {CV}. We choose the link {BRUS-ONSA}, i.e. Brussels-Onsala (Sweden), the day of an ionospheric storm, October 30, 2003, and used {GPS} observations with a satellite elevation cutoff of 5 degrees. \\
Figure \ref{Iono2Effect} shows the effect of applying the $I2$ corrections on {GPS} $P_{3}L_{3}$ analysis. 
We see an effect up to more than 10 picoseconds during the ionospheric storm on the link BRUS-ONSA. \\
The $I3$ effect shown in Figure \ref{Iono3Effect} is at the present noise level of GPS observations.
\\
Consequently, the residual ionospheric errors in $P_{3}L_{3}$ (when $P_{3}L_{3}$ is not corrected for higher-order effects) are mainly due to the contribution of $I2_{3}$.
A $I2$ delay of 15 picoseconds peak to peak during the storm (Figure \ref{2ndOrderIonoPerturbation}b) for a given station $A$ induces a variation with the corresponding differential $I2_{3A}$ - $I2_{3B}$ amplitude in {CV} frequency transfer with station $B$, as the shape of the curve is determined by the {GPS} phases for which the $I2$ correction is applied with a factor 1. Furthermore, $I2$ induces twice as much an offset on the absolute time synchronization error (Figure \ref{Iono2Effect}), as the calibration of the curve is determined by the code data for which the $I2$ correction is applied with a factor 2 (Equation \ref{eqGPSmeasurementsLP3}). However, this of course is still well below the present calibration capabilities of GPS equipment. \\
Note that the results presented here correspond to the time link {BRUS-ONSA}. It is therefore the differential ionospheric effect between those two stations 
that matters for the clock solution in Common-View mode. The impact of $I2$ on a clock solution in {PPP} could therefore be higher and induce larger effects on intercontinental time links. This will be investigated in further studies, when consistent satellite orbits and clock products computed with $I2$ (and $I3$) will be available. 

\section{Space weather}
\label{Section_Space_weather}

Solar activity was initially monitored thanks to ground observations (sunspot group evolution). 
Since recent decades, the Sun is observed via satellites. This opened the field to space weather studies,  characterizing the conditions of the Sun, of the space between the Sun and Earth, and on the Earth. \\
The Geostationary Operational Environmental Satellite ({GOES}, orbiting the Earth at 35790 km) measures solar flares as X-rays 
from 100 to 800 picometers 
and classifies them as A, B, C, M or X according to their peak flux (in watts per square meter) on a logarithmic scale. 
Each class has a peak flux ten times greater than the preceding one, with X ones of the order of $10^{-4}$ $W/m^{2}$. 
There is an additional linear scale from 1 to 9 inside each class. 

Furthermore, the magnetic field of the Earth is characterized (according to Earth latitude) by several indexes. 
The $K$ index quantifies disturbances in the horizontal component of the Earth's magnetic field with an integer in the range 0-9. 
$K$ index is derived from the maximum fluctuations of the Earth geomagnetic field horizontal component observed on a magnetometer 
during a three-hour interval. 
The conversion table from maximum fluctuation (in nanoTesla) to $K$-index, varies from observatory to observatory \citep{NOAA09}. 
Hence, the official planetary $K_{p}$ index is derived by calculating a weighted average of $K$-indices from a network of geomagnetic observatories. 
Values of $K$ from 5 and higher indicate a geomagnetic storm which may hamper {GPS}/{GNSS} TFT as we shall see.

\begin{figure}[p]
\begin{center}
\includegraphics*[width=20cm,angle=90]{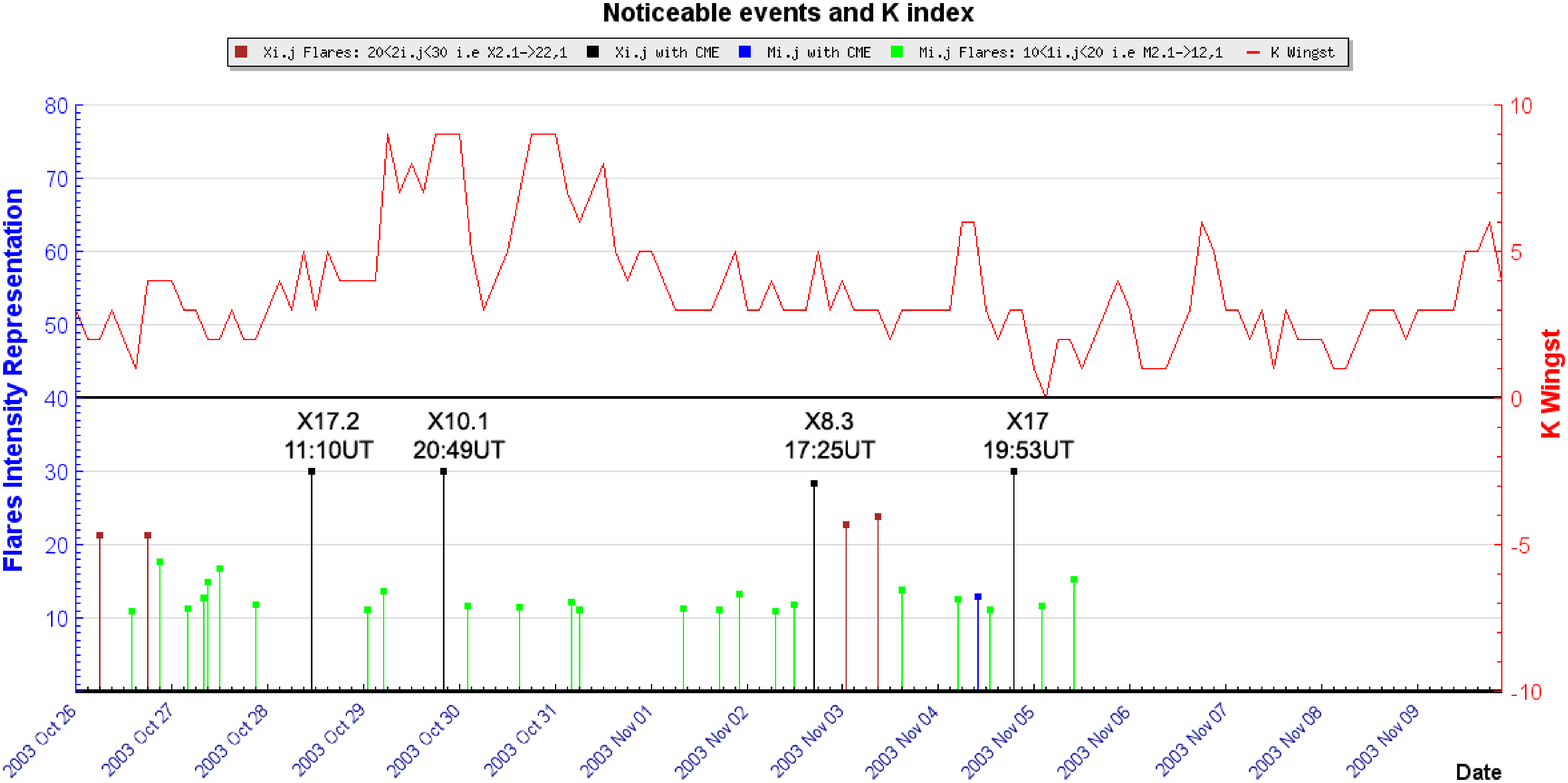}
\end{center}
\caption{Time series of the noticeable solar events (X and M X-ray flares) and geomagnetic $K$ index from Wingst Observatory ($K_{W}$ in the text). 
The notation used is the following. Labels $i$ and $j$ are linked to the classification of flares. M flares are represented by a line peaking between 10 and 20 (e.g. M2.5 by a line from 0 to 12.5), while X flares are represented between 20 and 30, at their respective peak-intensity time. 
Note that X flares bigger than X9.9 saturate the scale and are represented by a line peaking at 30.}
\label{HalloweenStorm}
\end{figure}

The previous solar maximum was around 2001. The year 2003 was thus an active period for the Sun, compared to year 2007. For the sake of comparison, we selected a quiet day, March 11 2007, and the very active day of 30th October 2003. Around our quiet test-day, no X and no M flares were observed; only four B-type X-ray events were detected during March 2007. On the other side, the solar activity and Earth geomagnetic conditions were at exceptionally high levels around the 30th October 2003, as can be seen from Figure \ref{HalloweenStorm}, which lists the solar flare events according to their intensity, together with the Earth geomagnetic index $K$ from Wingst Observatory ($K_{W}$ in the text) in Germany (at geographic latitude 53.74$^{\circ}$ and longitude 9.07$^{\circ}$, close to Onsala's latitude). Those events were mostly due to two large solar active regions named {NOAA}0486 and {NOAA}0484, which produced numerous solar M-flares and several X-flares. Further insight for Figure \ref{HalloweenStorm} can be gained when reading the corresponding {SIDC} (Solar Influences Data Center) weekly bulletins \citep{SIDC09} on solar activity for weeks 148 and 149 of 2003, corresponding to the range of our plots. We focus on the X-flares accompanied by {CME}.\\
  On October 28, a X17.2 (thus extremely strong) solar flare occurred from the solar active region {NOAA}0486 peaking at 11:10 UT. It was accompanied by a {CME} directed towards the Earth with an estimated plane-of-the-sky speed of about 2125 km/s, first detected at 10:54 UT in the {LASCO C2} field of view of the {SOHO} spacecraft (located at about $1.5\cdot10^{6}$ km from the Earth). As the CME was extremely fast, the shock arrived to the Earth around 06:00 UT on October 29. 
It produced a severe magnetic storm with $K_{W}$ index reaching 9. The intensity of the storm decreased slightly to $K_{W} = 7$ during October 29 as the arriving magnetic cloud was of the north-south type. The main portion of the negative the north-south interplanetary magnetic field component $B_{z}$ arrived in the trailing part of the cloud at the end of October 29. 
$B_{z}$ was strongly negative during around 8 hours producing the second peak in the $K_{W}$ index which reached 9 again during 18:00 - 24:00 UT. The $K_{W}$ index settled down to minor storm conditions ($K_{W}=5$) on October 30.\\
  On October 29, the solar active region {NOAA0486} produced an X10.1 flare peaking at 20:49 UT, as observed by {GOES}. It was associated with a {CME} observed by {SOHO/LASCO C2} at 20:54 UT that developed to a full halo {CME} directed towards the Earth with an estimated plane-of-the-sky speed of about 1950 km/s. The electromagnetic shock of the {CME} was registered by {ACE/MAG} around 16:00 UT on October 30. This time the magnetic cloud was of south-north type, so the severe geomagnetic storm started right after the arrival of the shock: the $K_{W}$ index reached 9 again and stayed at that level during two 3-hour intervals (18:00 - 24:00). The geomagnetic storm finally ended on October 31 - November 1 
when the $K_{W}$ index dropped to 4. \\
   From October 30 to November 5 included, several X- and M-class flares occurred in solar active regions {NOAA0484}, {NOAA0486} and {NOAA0488}. \\
  On November 2, an X8.3 flare was observed by {GOES}, peaking at 17:25 UT, from the solar active region {NOAA0486}. A halo {CME} with estimated plane-of-the-sky speed of about 2100 - 2200 km/s followed. The shock of the arriving {CME} was recorded in the solar wind on November 4, at 05:53UT. The associated interplanetary magnetic field pointed southward between 07:00 and 9:30 UT. This triggered a geomagnetic storm ($K_{W}$=6-7), but only for a limited duration: thereafter, the geomagnetic field remained quiet to unsettled.\\
  On November 4, came from {NOAA0486} an extreme X17 flare (peaking at 19:53UT), later estimated to have reached an X28 peak flux that saturated the {GOES detectors}, together with a full halo CME. The shock, corresponding to this CME which arrived sideways on November 6 at 19:37 UT, was relatively weak and only led to a minor storm that lasted until November 7, 00:00 UT. The magnetosphere then remained quiet to unsettled. \\
  By contrast, during the end of the week, the Sun had a very low activity. 
However, an existing large solar coronal hole 
gave rise to minor geomagnetic storm ($K_{W}=6$) episodes on November 9.
  
\begin{figure}[p]
\begin{center}
\includegraphics*[width=15cm,angle=-90]{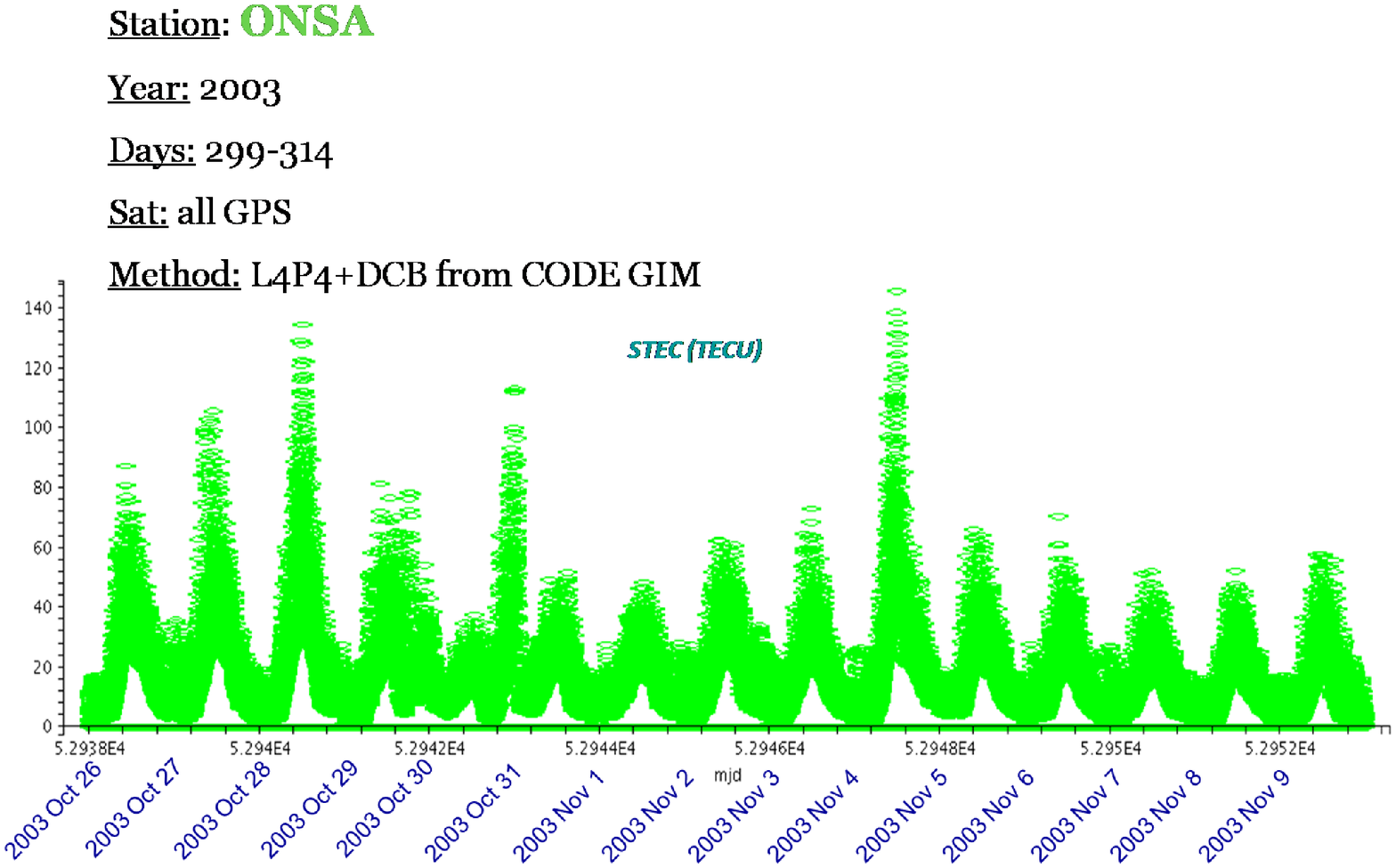}
\end{center}
\caption{STEC time series computed with the {ATOMIUM} software for station Onsala (ONSA), according to Equation \ref{eqSTECarc}, from 26th October to 9th November 2003.}
\label{STECHalloweenStorm}
\end{figure}

\begin{figure}[p]
\begin{center}
\includegraphics*[width=15cm,angle=0]{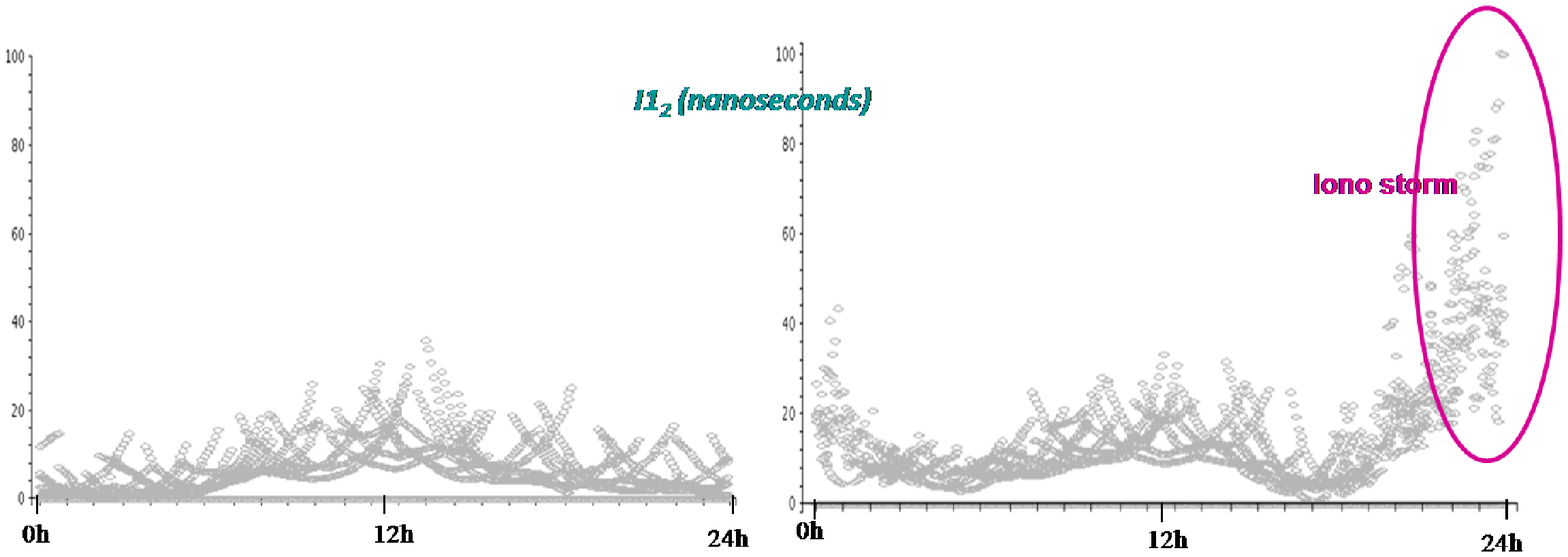}
\end{center}
\caption{First-order ionospheric delay in GPS frequency 2, for station Onsala on an ionosphere-quiet day, 11th March 2007 (Figure a, left), versus on an ionosphere-stormy day, 30th October 2003 (Figure b, right).}
\label{1stOrderIonoPerturbation}
\end{figure}

\begin{figure}[p]
\begin{center}
\includegraphics*[width=15cm,angle=0]{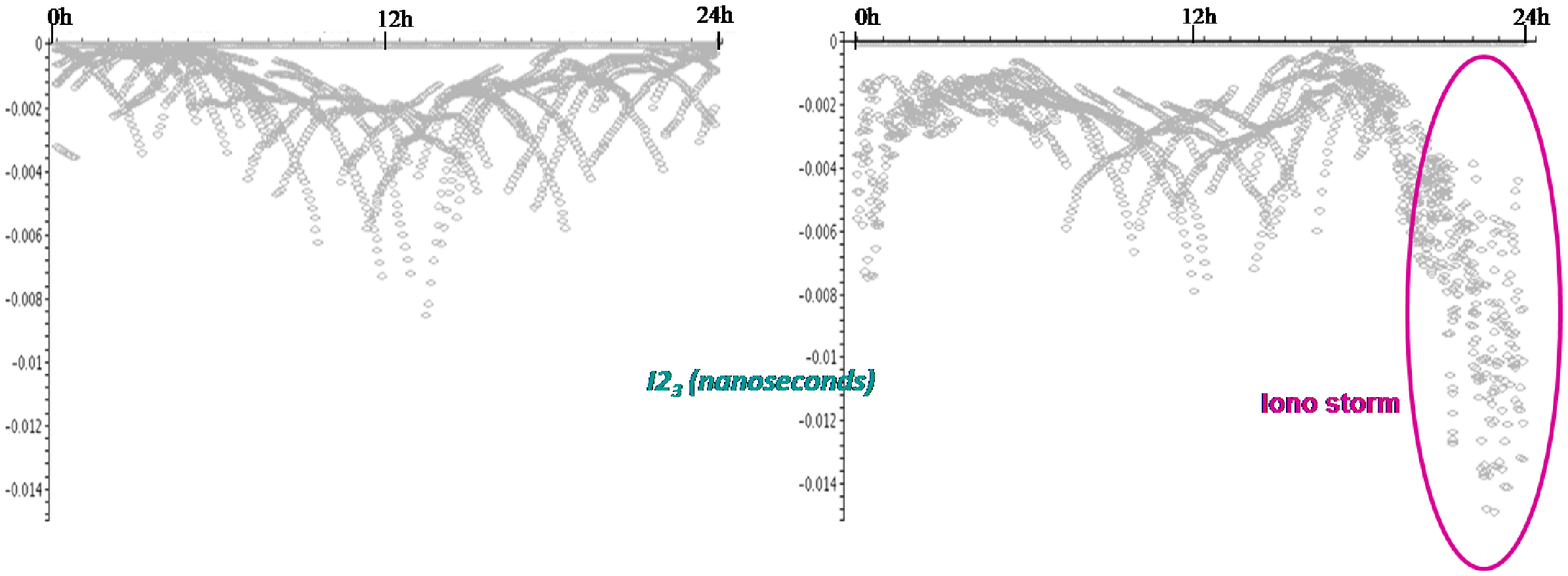}
\end{center}
\caption{Second-order ionospheric delay in GPS so-called ionosphere-free combination, for station Onsala on an ionosphere-quiet day, 11th March 2007 (Figure a, left), versus on an ionosphere-stormy day, 30th October 2003 (Figure b, right).}
\label{2ndOrderIonoPerturbation}
\end{figure}

\begin{figure}[p]
\begin{center}
\includegraphics*[width=15cm,angle=0]{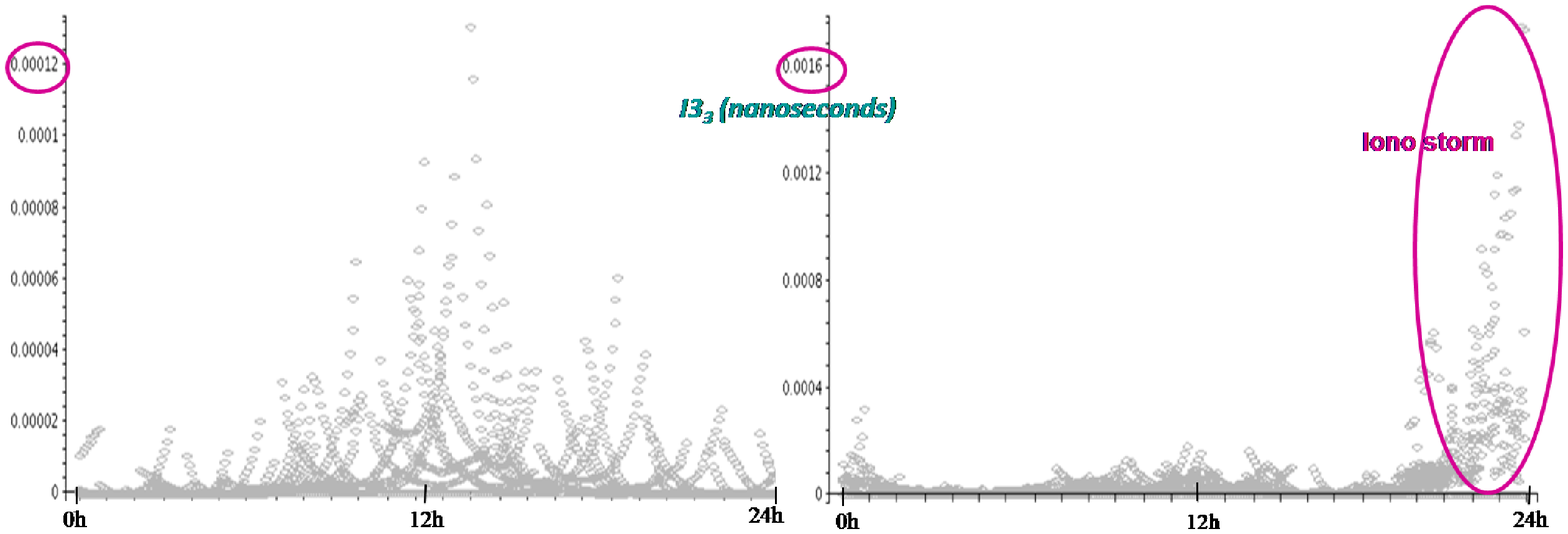}
\end{center}
\caption{Third-order ionospheric delay in GPS so-called ionosphere-free combination, for station Onsala on an ionosphere-quiet day, 11th March 2007 (Figure a, left), versus on an ionosphere-stormy day, 30th October 2003 (Figure b, right).}
\label{3rdOrderIonoPerturbation}
\end{figure}

\begin{figure}[p]
\begin{center}
\includegraphics*[width=13cm,angle=0]{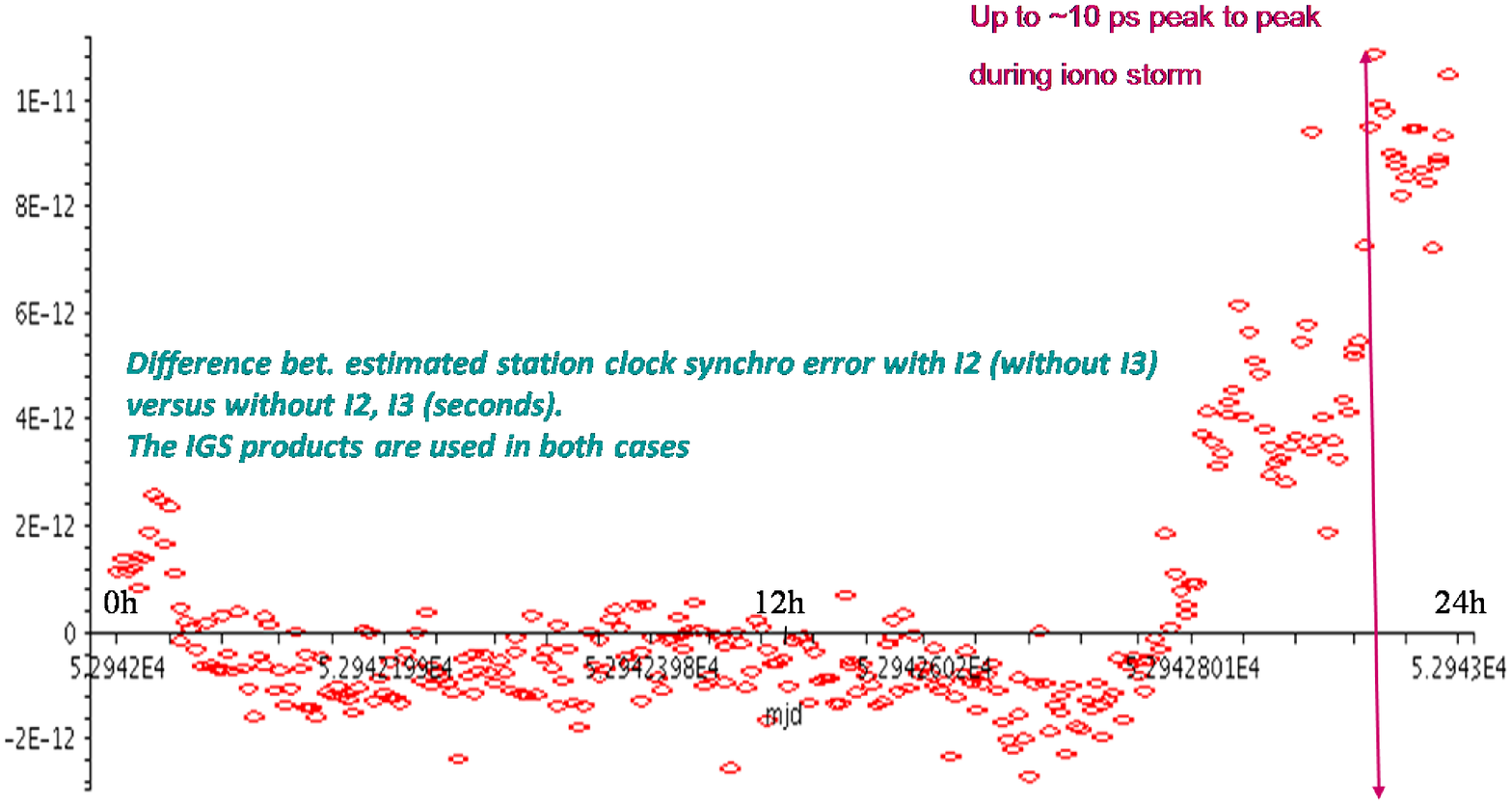}
\end{center}
\caption{Effect of taking second-order ionospheric effects, or not, into account in the $L_{3}P_{3}$ GPS measurements for the Brussels-Onsala link, on the ionosphere-stormy day 30th October 2003. The difference is taken between two ATOMIUM estimated station clock solutions, both using IGS products.}
\label{Iono2Effect}
\end{figure}

\begin{figure}[p]
\begin{center}
\includegraphics*[width=13cm,angle=0]{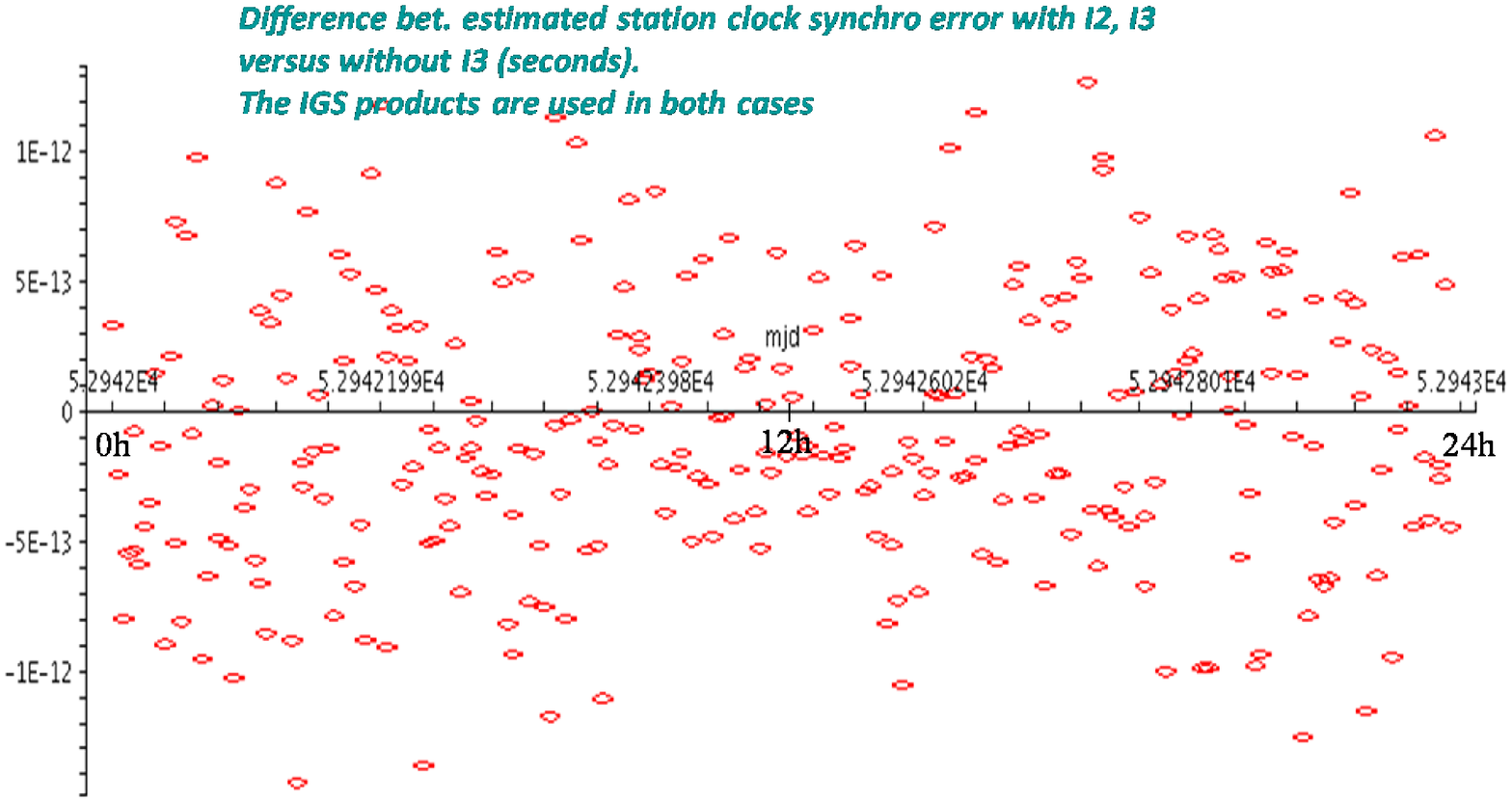}
\end{center}
\caption{Effect of taking third-order ionospheric effect, or not, into account in the $L_{3}P_{3}$ GPS measurements for the Brussels-Onsala link, on the ionosphere-stormy day 30th October 2003. The difference is taken between two ATOMIUM estimated station clock solutions, both using IGS products.}
\label{Iono3Effect}
\end{figure}

We now come back to our {STEC} at station Onsala on the 30th October 2003 (Figure \ref{STECHalloweenStorm}), estimated using the software {ATOMIUM} and {GPS} data. We understand that the huge solar flare event X17.2 of October 28 2003, that was associated with a {CME}, through the Earth geomagnetic storm it triggered on the 29th October 2003, is responsible for the abnormally high level of {STEC} at Onsala (and Brussels as well) at early hours of October 30 2003. This explains the correspondingly high first-, second- and third-order ionospheric perturbations computed in Figures \ref{1stOrderIonoPerturbation}, \ref{2ndOrderIonoPerturbation} and \ref{3rdOrderIonoPerturbation}, in the first hours of October 30 2003. A small corresponding impact can be seen on the {ATOMIUM} clock solution in Figure \ref{Iono2Effect}. \\
The major Earth geomagnetic storm that occurred on the 30th October 2003, however, is most probably due to the solar flare event X10.1 of October 29 2003, associated with a {CME}. This time, the strong signature present in the {STEC} estimated with {ATOMIUM} for Onsala (Figure \ref{STECHalloweenStorm}) (and Brussels as well) and computed corresponding first-, second- and third-order ionospheric perturbations on {GPS} signals (Figures \ref{1stOrderIonoPerturbation}, \ref{2ndOrderIonoPerturbation} and \ref{3rdOrderIonoPerturbation}) is clearly visible in the estimated Common-View clock solution of {ATOMIUM} (Figure \ref{Iono2Effect}). This emphasizes the need for higher-order ionospheric corrections on {GPS} signals in precise time and frequency transfer, as we need to prepare for the next solar maximum.

\newpage
\section{Conclusions}
\label{Section_Conclusions}

This study aimed at quantifying the impact of second- and third-order ionospheric delays on precise time and frequency transfer using GPS signals.

We used the {ATOMIUM} software, in which we implemented higher-order ionospheric contributions (second and third orders) in the so-called ionosphere-free combination of {GPS} codes and phases. 
We then compared these ionospheric contributions with the first-order ionospheric effect on the {GPS} dual frequency signal, which is cancelled in ionosphere-free combination.
It was shown that the first-order ionospheric delay of several tens of nanoseconds on an ionosphere-quiet day, is doubled in case of ionospheric storms. Though second-order delays in the ionosphere-free combination are about 3 to 4 orders of magnitude smaller than the first-order delays, they can reach about 15 picoseconds on a stormy day, which is significant when performing geodetic time and frequency transfer with very stable clocks. Third-order delays in the ionosphere-free combination are yet an order of magnitude smaller, and are at the level of present noise of {GPS} observations.
The impact of those higher-order delays on ionosphere-free time and frequency transfer clock solutions was estimated for the time link {BRUS-ONSA}. It reaches more than 10 picoseconds during the ionospheric storm of October 30 2003.
Finally, we illustrated the correlations between space weather data (strong solar events, disturbances of the Earth geomagnetic field) and time/frequency transfer performed with {GPS} signals.

\newpage

\newpage

\ack{This work has been supported by the Solar and Terrestrial Center of Excellence \citep{STEC09}. The authors also acknowledge the IGS for their data and products \citep{IGSa09}, 
and the Solar Influences Data Analysis Center (SIDC) \citep{SIDC09} for their space weather weekly bulletin products and for their archives on solar events and related indexes.}\bigskip


\clearpage
\begin{table}
\caption{Orders of magnitude of ionospheric effects $I1$, $I2$ and $I3$ on GPS phase measurements 
(for codes, see converting factor in code measurement Equations \ref{eqGPSmeasurementsLP12}, \ref{eqGPSmeasurementsLP3}, Section 1)}
\begin{tabular}{c}
\begin{tabular}{llll}
\hline
Ionospheric effect & Delay in  $L_{1}L_{2}$ & Delay in  $L_{3}$ & Relevance \\
(absolute value) & per 100 TECU & per 100 TECU & \citep{H08}\\
\hline
$I1$ & $\sim 30 ns - 100$ ns & $0$ & 99.9\%  of $I123…$\\
$I2$ & $\sim 0 - 130$ ps & $\sim 0 - 45$ ps & 90\%  of $I23$\\
$I3$ & $\sim 0 - 3$ ps & $\sim 0 -  2$ ps & \\
\hline\
\end{tabular}
\end{tabular}
\label{table}
\end{table}


\begin{thebibliography}{}


\bibitem[Dach et al.(2007)]{DHFM07}
Dach, R., Hugentobler, U., Fridez, P., \& Meindl, M., 
{\em Bernese GPS software version 5.0}, 1-612, 2007.

\bibitem[Defraigne et al.(2008)]{DGB08}
Defraigne, P., Guyennon, N., \& Bruyninx, C., 
GPS Time and Frequency Transfer: PPP and Phase-Only Analysis, 
International Journal of Navigation and Observation, 
Vol. 2008, Article ID 175468, 1-7, 2008.

\bibitem[Bassiri \& Hajj(1993)]{BH93}
Bassiri, S. , \& Hajj, G. A., 
Higher-order ionospheric effects on the global positioning system observables and means of modeling them, 
Manuscripta Geodetica 18, 280-289, 1993.

\bibitem[Berghmans et al.(2005)]{B05}
Berghmans, D., Van der Linden, R.A.M., Vanlommel, P., Warnant, R., Zhukov, A.N.,Robbrecht, E., Clette, F., Podladchikova, O., Nicula, B.,  Hochedez, J.-F., Wauters, L., \& Willems, S.,
Solar activity: nowcasting and forcasting at SIDC,
Annales Geophysicae, 23(9), 3115-3128, 2005.

\bibitem[Brunner \& M. Gu(1991)]{BG91}	
Brunner, F. K., \& Gu, M., 
An improved model for the dual frequency ionospheric correction of GPS observations, 
Manuscripta Geodetica, 16, 205-214, 1991.

\bibitem[Fritsche et al.(2005)]{FDKRV05}
Fritsche, M. , Dietrich, R., Kn\"{o}fel, C., R\"{u}lke, A., \& Vey, S., 
Impact of higher-order ionospheric terms on GPS estimates, 
Geophysical Research Letters 32, L23311, 1-5, Formula (14), 2005.

\bibitem[Hernandez-Pajares et al.(2007)]{HJSO07}
Hernandez-Pajares, M., Juan, J. M., Sanz, J., \& Oruz, R.,
Second-order ionospheric term in GPS: Implementation and impact on geodetic estimates, 
Journal of Geophysical Research, 112, B08417, 1-16, 2007.

\bibitem[Hernandez-Pajares et al.(2008)]{H08}
Hernandez-Pajares, M., Fritsche, M., Hoque, M.M., Jakowski, N., Juan, J.M., Kedar, S., 
Krankowski, A., Petrie, E., \& Sanz, J., 
Methods and other considerations to correct for higher-order ionospheric delay terms in GNSS, 
IGS workshop, Miami, USA, 2008, oral presentation.

\bibitem[Hoque \& Jakowski(2008)]{HJ08}
Hoque, M. M., \& Jakowski, N., 
Estimate of higher order ionospheric errors in GNSS positioning, 
Radio Science, 43, 1-15, RS5008, doi:10.1029/2007RS003817, 2008. 

\bibitem[IGS atx(2009)]{IGSb09}
IGS satellite antenna phase center variations files igs01.atx and igs05.atx,
ftp://igscb.jpl.nasa.gov/pub/station/general/ 
%
%
%
\bibitem[IGS data and products(2009)]{IGSa09}
IGS data and products. ftp://igscb.jpl.nasa.gov/

\bibitem[Jakowski et al.(1994)]{JPM94}
Jakowski, N., Porsch, F., \& Mayer, G., 
Ionosphere - Induced -Ray-Path Bending Effects in Precision Satellite Positioning Systems, 
SPN 1/94, 6-13, 1994

\bibitem[Kouba \& Heroux(2001)]{KH01}
Kouba, J., \& Heroux, P., 
GPS Precise Point Positioning using GPS orbit products, 
GPS solutions, 5, 12-28, 2001

\bibitem[Leick(2004)]{L04}
Leick, A., {\em GPS satellite surveying}, 
3rd Edition, John Wiley and Sons INC, 2004

\bibitem[Lyard et al.(2004)]{LLL04}
Lyard, F., Lefevre, F., Letellier, T., \& Francis, O, 
Modelling the global ocean tides: modern insights from FES2004, 
Ocean Dynamics, 56, 394–415, 2006

\bibitem[McCarthy \& Petit(2003)]{MP03}
McCarthy, D., \& Petit, G.,
{\em IERS Conventions 2003: IERS Technical Note 32}, 
Frankfurt am Main: Verlag des Bundesamts f\"{u}r Kartographie und Geod\"{a}sie, 1-127,2004, ISBN 3-89888-884-3, http://www.iers.org/documents/publications/tn/tn32/tn32.pdf

\bibitem[Niell(1996)]{N96}
Niell, A. E.,
Global mapping functions for the atmospheric delay at radio wavelengths, 
Journal of Geophysical Research, 101(B2), 3227-3246,1996.

\bibitem[NOAA Space Weather Prediction Center(2009)]{NOAA09}
NOAA Space Weather Prediction Center, the K index,
http://www.swpc.noaa.gov/info/Kindex.html

\bibitem[Saastamoinen(1972)]{S72}
Saastamoinen, J.,
Atmospheric corrections for the troposphere and stratosphere in radio ranging of satellites, 
Geophysical Monograph 15, Use of Artificial Satellites for Geodesy, 247-251, AGU 1972, 1972.

\bibitem[SIDC(2009)]{SIDC09}
Solar Influences Data Analysis Center (SIDC), http://sidc.oma.be/, 
hosted by the Royal Observatory of Belgium;
SIDC space weather weekly bulletin: http://www.sidc.be/products/bul/index.php, bulletins 307 and 314.

\bibitem[STCE(2009)]{STEC09}
Solar- Terrestrial Center of Excellence (STCE), http://www.stce.be/index.php 
%
%
\bibitem[Tsyganenko(2005)]{T05}
Tsyganenko, N.A., 
A set of FORTRAN subroutines for computations of the geomagnetic field in the Earth's magnetosphere, 
version of May 4, 2005, 
available on http://modelweb.gsfc.nasa.gov/magnetos/tsygan.html, as geopack-2005.doc and full fortran routines.

\bibitem[Wu et al.(1993)]{WWHBL93}
Wu, J.T. , Wu, S. C., Hajj, G.A., Bertiger, W.I., \& Lichten, S.M., 
Effects of antenna orientation on GPS carrier phase, 
Manuscripta Geodetica 18, 91-98, 1993.

\end{thebibliography}
\end{document}